\documentclass[preprint]{emulateapj} 

\shorttitle{Initial Conditions of the Nice Model} 
\shortauthors{Batygin \& Brown} 

\begin{document}
 
\title{Early Dynamical Evolution of the Solar System:  \\
Pinning Down the Initial Condition of the Nice Model}  

\author{Konstantin Batygin \& Michael E. Brown } 

\affil{Division of Geological and Planetary Sciences, California Institute of Technology, Pasadena, CA 91125}

\slugcomment{Accepted to The Astrophysical Journal}
\email{kbatygin@gps.caltech.edu}
 
\begin{abstract} 
In the recent years, the ``Nice" model of solar system formation has attained an unprecedented level of success in reproducing much of the observed orbital architecture of the solar system by evolving the planets to their current locations from a more compact configuration. Within the context of this model, the formation of the classical Kuiper belt requires a phase during which the ice giants have a high eccentricity. An outstanding question of this model is the initial configuration from which the Solar System started out. Recent work has shown that multi-resonant initial conditions can serve as good candidates, as they naturally prevent vigorous type-II migration. In this paper, we use analytical arguments, as well as self-consistent numerical N-body simulations to identify fully-resonant initial conditions, whose dynamical evolution is characterized by an eccentric phase of the ice-giants, as well as planetary scattering. We find a total of eight such initial conditions. Four of these primordial states are compatible with the canonical "Nice" model, while the others imply slightly different evolutions. The results presented here should prove useful in further development of a comprehensive model for solar system formation.

\end{abstract}


\section{Introduction} 

The question of how the solar system formed dates back centuries. The last decade, however, has seen a considerable amount of progress made on this issue. Notably, the development of the Nice model (Tsiaganis et al. 2005) has proven to be a milestone. The scenario foretold by the Nice model is as follows: the giant planets form in a compact configuration, and driven by planetesimal scattering (Fernandez \& Ip 1984, Malhotra 1995) begin migrating divergently. Eventually, Jupiter and Saturn cross their mutual 2:1 mean-motion resonance (MMR), which results in an acquisition of eccentricities for both planets. Subsequently, the whole outer solar system undergoes a brief period of dynamical instability, during which Uranus and Neptune are scattered to their current orbits. 

There are a few aspects to the success of the Nice model. First and foremost, it has been able to replicate the architecture of the secular dynamics of the outer solar system (Morbidelli et al. 2009a). Second, it provides a semi-quantitative description of the formation of the Kuiper belt (Levison et al. 2008). Third, the inward flux of planetecimals during the phase of dynamical instability allows for chaotic capture of Jupiter's and Neptune's Trojan populations (Morbidelli et al. 2005, Nesvorni{\'y} et al. 2007). Finally, if the resonance crossing between Jupiter and Saturn is timed appropriately, the global mayhem provides a natural trigger for Late Heavy Bombardement (LHB) (Gomes et al. 2005). There are other observational constraints that should be reproduced in a model for the solar system's formation, such as the dynamical structure of the inner solar system, and considerable progress has already been made in this direction (Brasser et al. 2009, Morbidelli et al. 2009b). At the same time, it is also crucial to explore the unobservable aspect of the model, namely the initial conditions. This is the purpose of our study.

In this paper, we consider various multi-resonant configurations as possible initial conditions for the Nice model. We investigate the early stages of dynamical evolution for a large number of candidate systems and show that only eight configurations appear to be consistent with the formation of the Kuiper belt in the framework of the Nice model. It is noteworthy that running simulations of the Nice model to completion is computationally expensive. Consequently, as we seek to examine a large array of initial conditions, we are forced to utilize early dynamical events, namely planetary scattering, as proxies for successful formation of the solar system. In this manner, we limit the duration of each simulation to only a few tens of millions of years. Within the context of our integrations, this is long enough for the system to pass through the epoch of dynamical instability, but not long enough to scatter away all of the planetesimals which end up on long-term unstable orbits.

The plan of the paper is as follows: in section 2, we explain how multi-resonant configurations prevent type-II migration, and our approach to their assembly. In section 3, we discuss the evolution scenarios of the considered systems, estimate the amplitudes of eccentricity jumps in relevant cases, and present the results of N-body simulations. We discuss our results and conclude in section 4.

\section{Multi-Resonant Configurations}

One of the important differences between the solar system and the majority of the detected aggregate of extra-solar planetary systems is the lack of a close-in giant planet. This difference suggests that while it is common for planets to migrate to small orbital radii, some mechanism was at play in the early solar system which prevented vigorous orbital decay. One such mechanism, which is both efficient and reasonable, is resonant capture (Masset  \& Snellgrove 2001). 

When a newly formed gaseous planet reaches a critical mass of $\sim 1 M_J$, it opens a gap in the proto-planeatary disk. Incidentally, the planet continues to interact with the disk via various resonances. Summed together, the resonant torques from a given side of the disk, somewhat counter-intuitively, push the planet away from that side. As a result, the planet positions itself at a point in the gap where all torques cancel, and moves inward together with the disk on the viscous time-scale, in a process termed Type-II migration (Morbidelli \& Crida 2007). Simultaneously if another planet, whose semi-major axis is larger, is migrating inwards faster, it will eventually encounter a mean-motion resonance. Under a large spectrum of circumstances, converging orbits can lead to capture into a mean-motion resonance, ensuring that the two planets' period ratio remains constant for extended periods of time. In fact, for slow enough migration rates and low enough eccentricities, resonant capture is certain (Peale 1986). When this happens, the gas between the two planets drains as the gaps overlap. Consequently, the torque balance on the resonant pair results from gas interior to the inner planet and that exterior to the outer planet. This leads to a drastic reduction of the migration rate (Lee \& Peale 2002). Furthermore, if the inner planet is more massive than the outer planet, as is the case with Jupiter and Saturn, the migration of the resonant pair can be halted altogether or even reversed (Morbidelli \& Crida 2007). 

Numerical studies of Jupiter and Saturn submerged in a gaseous proto-planetary disk suggest precisely the above scenario. Indeed, Saturn's migration is considerably faster than Jupiter's because of its lower mass and inability to fully open a clean gap. The pioneering results of Masset  \& Snellgrove (2001) showed that locking Jupiter and Saturn in the 3:2 MMR can effectively halt the pair's migration. The somewhat more precise numerical experiments of Morbidelli \& Crida (2007) confirmed this and also showed that capture into 2:1 and 5:3\footnote{It must be noted that capture into 5:3 MMR is less probable, since it is a second order resonance. Furthermore, even if Jupiter and Saturn are captured, subsequent motion can be unstable (Morbidelli \& Crida 2007). } MMR's are viable outcomes, depending on where Saturn forms relative to Jupiter. The work of Pierens \& Nelson (2008) however suggests that while capture into 2:1 and 5:3 MMR's is certainly possible, in a number of cases Saturn eventually breaks away and continues its inward migration until it is captured in the 3:2 MMR. Collectively, the above mentioned results suggest that 3:2 MMR is indeed a likely initial configuration of Jupiter \& Saturn, although there is not enough evidence to decisively rule out the 2:1 MMR or the 5:3 MMR as initial conditions. Consequently, for the sake of completeness, we consider all three of these resonances as possible starting configuratoins for Jupiter and Saturn.

By extension of the above scenario, the ice giants, which are believed to form after Jupiter and Saturn, behave in a qualitatively similar way. Namely, as they migrate from the outer disk inwards, they too become trapped in MMR's. Consequently, at the epoch of the disappearance of the gas, we are left with a multi-resonant system, in which each planet is in resonance with its neighbors. Morbidelli et al. (2007) performed hydrodynamical simulations of this process with the purpose of identifying such configurations that are long-term stable. Considering only systems where Jupiter and Saturn are locked in a 3:2 MMR, they were able to find two stable fully-resonant states.

Our approach to assembling multi-resonant systems follows that of Lee \& Peale (2002). In addition to Newtonian N-body interactions, each planet is subject to semi-major axis decay
\begin{equation}
\frac{\dot{a}}{a} = K,
\end{equation}
and eccentricity damping
\begin{equation}
\frac{\dot{e}}{e} = 10^2 \left( \frac{\dot{a}}{a} \right),
\end{equation}
where $a$ is semi-major axis, $e$ is the eccentricity, and $K$ is an adjustable migration frequency. In our simulations, we keep K the same for all planets, ensuring always convergent migration. In accord with Lee \& Peale (2002), a Bulirsch-Stoer integration method (Press et al. 1992) was used. In contrast with the full hydrodynamical simulations, this method is simpler and computationally cheaper, allowing us to sample a large array of systems. Additionally, given the problem's straight-forward nature, it is unlikely that a configuration found using this approach cannot be obtained using other methods.

In the context of these simulations, all four giant planets were introduced simultaneously on planar circular orbits, slightly outside of their desired resonant locations, with the more massive ice giant on the outermost orbit. If capture into the desired resonances did not occur, we varied $K$. As pointed out in Morbidelli et al. (2007), the sequence in which planets get captured may be important, since changing the order can change the librating resonant angles. To avoid this degeneracy, we always set the initial orbits such that Saturn would get captured first and Neptune last. To avoid confusion, we shall always refer to the outermost planet as Neptune, although in the Nice-model, the orbits of the ice giants may switch places. After each desired configuration was achieved, the 5-body system was subjected to a 100Myr dynamical stability test, using the mercury6 software (Chambers 1999). Note that 100Myr is highly conservative, given that our full dynamical evolution simulations only last $\sim$30Myr. However, these stability integrations show that the same initial conditions are also applicable for scenarios where the global instability occurs somewhat later than what is considered in this work. Table (1) lists all stable multi-resonant configurations that we generated.

\begin{table}
\begin{center}
\caption{Multi-resonant  Initial Conditions}
\begin{tabular}{cccccccccc}
\\
\tableline\tableline
J:S & S:U  & U:N\\
\tableline
3:2&2:1&3:2\\
3:2&2:1&4:3\\
3:2&3:2&3:2\\
3:2&3:2&4:3\\
\textbf{3:2}&\textbf{3:2}&\textbf{5:4}\\
\textbf{3:2}&\textbf{4:3}&\textbf{3:2}\\
\textbf{3:2}&\textbf{4:3}&\textbf{4:3}\\
\\
5:3&2:1&3:2\\
5:3&2:1&4:3\\
5:3&2:1&5:4\\
5:3&2:1&6:5\\
5:3&3:2&3:2\\
5:3&3:2&4:3\\
\textbf{5:3}&\textbf{3:2}&\textbf{5:4}\\
\textbf{5:3}&\textbf{3:2}&\textbf{6:5}\\
\textbf{5:3}&\textbf{4:3}&\textbf{3:2}\\
\textbf{5:3}&\textbf{4:3}&\textbf{4:3}\\
5:3&4:3&5:4\\
\\
2:1&2:1&3:2\\
2:1&2:1&4:3\\
2:1&3:2&3:2\\
2:1&3:2&4:3\\
2:1&4:3&3:2\\
\textbf{2:1}&\textbf{4:3}&\textbf{4:3}\\

\tableline
\end{tabular}
\end{center}
\tablenotetext{0}{All of the stable multi-resonant initial conditions considered in this study. The bold lines represent the configurations that proved to be compatible with a Nice model-like evolution.}
\end{table}

\section{Dynamical Evolution}

Having found a large array of stable multi-resonant systems, we now need to determine which of these configurations can resemble the current state of the solar system, having dynamically evolved. There are two constraints of interest here. First is the structure of the secular dynamics of the giant planets. Recently, Morbidelli et al. (2009a) showed that a smooth migration scenario, such as the one envisioned by Malhotra (1995), is incompatible with the observed eccentricities and inclinations of the giant planets. A mean motion resonance crossing event by itself is also insufficient because it does not excite the inclinations to a necessary degree or reproduce the amplitudes of the $g_5$ and $g_6$ secular eigenmodes correctly. To create the current eccentricities, inclinations, and eigenmode amplitudes, encounters must have happened between an ice giant and a gas giant. Furthermore, if the instability took place after the inner solar system was already intact, encounters must have taken place between an ice giant and both gas giants to cause Jupiter to "jump", in order to prevent slow secular resonance sweeping of the inner solar system (Brasser et al. 2009). 

The second criterion of interest is the formation of the Kuiper belt, particularly the classical region. The transport mechanism of planetesimals to this region, proposed by Levison et al. (2008) relies on overlapping mean motion resonances. Namely, when Neptune's eccentricity exceeds $\sim0.2$, its exterior MMR's widen enough to overlap, and motion of all particles in the region becomes highly chaotic. This allows for planetesimals to execute a random-walk and invade the classical Kuiper belt region. With time, as Neptune's eccentricity decays due to dynamical friction (Stewart \& Withrill 1988), the resonances become narrower, and the particles occupying the classical region, no longer chaotic, remain on their orbits forever.

Using the two constraints described above as proxies for a successful formation scenario, we look for a subset of our generated initial conditions that result in evolutions which encompass both a scattering event and a transient high-eccentricity phase of the outer ice giant. In the Nice model, these two constraints are practically always satisfied simultaneously, since the ice giants tend to switch places and the scattering event is intimately tied to the high eccentricities. 

Our dynamical evolution simulations include the four outer planets and a disk of $\sim 3000$ equal-mass planetesimals, while the mass of the inner solar system is added to the sun. The radial surface density of the planetesimal swarm was assumed to have a power law structure: $\Sigma \propto r^{-k}$, where $ k \in (1,2)$. Consequently, the set of initial-value problems at hand is controlled by three parameters: Jupiter's semi-major axis, which due to  resonant relations controls the semi-major axes of the other planets, the planetesimal disk's mass, $m_{disk}$, and $k$. As it turns out, the actual value of $k$ has little effect on whether a given initial condition gives rise to a scattering event and an eccentric Neptune. Rather, it controls how fast the planets grind through the disk. As a result, we allow it to float randomly from simulation to simulation. Also, an advantage of multi-resonant initial conditions lies in that with enough simulations, it is possible to determine a unique combination of $(a_J,m_{disk})$, since Jupiter and Saturn must scatter enough particles to arrive to their current 5:2 commensurability.

In our simulations, the inner edge of the planetesimal disk was placed $\sim$ 1AU outside of Neptune's orbit. This forces migration, driven by planetesimal scattering, to begin shortly after the start of the simulation. In other words, we do not attempt to time the onset of instability with LHB, as was done in Gomes et al. (2005). There is another implication of a relatively close inner edge. If the planetesimal swarm is nearby when the instability begins, the planetary orbits penetrate deeply into the disk, and the resulting dynamical friction plays a stabilizing effect (Levison, personal communication). Intuitively, it makes sense to place the inner edge of the disk where the dynamical lifetime of planetesimals equals the lifetime of the gaseous nebula. While a 1 AU separation is approximately correct (Gomes et al. 2005), in the future it may be a worthwhile exercise to determine this boundary precisely for each multi-resonant initial condition. The outer edge of the disk was placed at 30 AU with the purpose of eventually halting Neptune's migration (Tsiaganis et al. 2005, Levison et al. 2008).

In order to not force the scattering-driven migration fictitiously, but still keep the computational cost down, we ignore self-gravity of the planetesimal disk. The resulting Hamiltonian takes the form:
\begin{equation}
H=\sum_{i=1}^{N}\frac{p_i^2}{2m_i} -G\sum_{i=1}^{5}m_{i}\sum_{j=i+1}^{N}\frac{m_{j}(x_i-x_j)}{|x_i-x_j|^2}
\end{equation}
where $p$ is momentum, $x$ is position, $m$ is mass and $G$ is the universal gravitational constant. To further diminish the computational cost, after a particle was scattered beyond 500AU, it was removed from the simulation. It must be noted that planetesimal-planetesimal interactions may in reality be important, since they give rise to an effective viscosity in the swarm (Levison, personal communication). Thus, more careful validation of our results should preferentially include these thorny effects. We used a hybrid Bulisch-Stoer/Wisdom-Holman algorithm of the mercury6 software for all integrations. We consistently used a time-step of $\tau = 300$d, and checked all successful simulations with a smaller $\tau = 60$d time-step to ensure that the observed instability is not a numerical artifact. In all such checks, the evolutions were practically indistinguishable, which assures that integrals of motion are sufficiently conserved. Finally, our simulations only cover a few tens of millions of years, since the focus here is on distinguishing between initial conditions that give rise to gas-giant/ice-giant scattering and ones that don't. As a result, the long-term evolution of the system after the instability is unexplored.

\subsection{Initial Conditions with Jupiter and Saturn in a 3:2 MMR}

Let us first consider a family of initial conditions, listed in table (1), where Jupiter and Saturn are in a 3:2 MMR.  This family of initial conditions was previously studied in some detail by Morbidelli et al. (2007).  Consequently, this section's results are partially reproductions. Using a hydrodynamical model, Morbidelli et al. (2007) found six multi-resonant configurations, two of which they determined to be long-term stable. These are the configurations listed in table (1) where both Jupiter \& Saturn and Saturn \& Uranus pairs are in 3:2 MMR's while Uranus \& Neptune are in either 4:3 or 5:4 MMR's. There are two more compact configurations listed in table (1) which we determined to be stable, although the counterparts of these configurations put together by Morbidelli et al. (2007) were unstable\footnote{Note that our simplified migration model is computationally cheaper than that of Morbidelli et al. (2007), allowing us to run more trials to arrive at stable configurations.}. These configurations are the two where Jupiter \& Saturn are in a 3:2 MMR, Saturn \& Uranus are in a 4:3 MMR, and Uranus \& Neptune are either in a 3:2 MMR or 4:3 MMR. 

As discussed in Morbidelli et al. (2007), if Jupiter and Saturn start out in a 3:2 MMR, the instability is triggered by their encounter with the 5:3 MMR. While this is a second-order resonance, small jumps in Jupiter's and Saturn's eccentricities go a long way, especially in highly compact configurations. Unfortunately, in this case it is difficult to conclusively determine which configurations will result in evolutions with scattering events a-priori. Thus, we must rely solely on numerical integrations to explore the various evolutionary outcomes of these initial conditions. 

After an initial run of 20 integrations for each initial condition of the family listed in table (1), we ruled out the configurations where Saturn \& Uranus are in a 2:1 MMR as well as the configuration where all planet pairs are in 3:2 MMR's because all evolutions were characterized by smooth migration. We subjected the remaining four configurations to 30 additional integrations and found that the only  configuration which does not result in ice-giant/gas-giant scattering is the one where Saturn \& Uranus are in a 3:2 MMR and Uranus \& Neptune are in a 4:3 MMR. The evolutions of the remaining initial conditions are presented in figures (1) - (3), and their final orbital parameters are entered into table (2).

For the initial condition in which Saturn and Uranus are in a 3:2 MMR, 10\% of the integrations were successful with 57\% of them exhibiting close encounters between an ice giant and \textit{both} gas giants. The same fractions for the two configurations where Saturn and Uranus are initially in a 4:3 MMR are 20\% \& 30\% and 27\% \& 50\% for the case where Uranus \$ Neptune are in a 3:2 MMR and 4:3 MMR respectively.

\begin{figure}[t]
\includegraphics[width=0.5\textwidth]{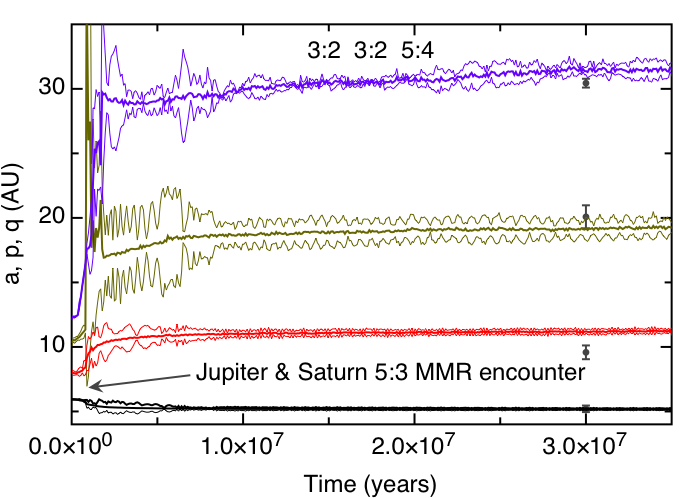}
\caption{Dynamical evolution of the initial configuration where initially Jupiter \& Saturn are in a 3:2 MMR, Saturn \& Uranus are in a 3:2 MMR and Uranus \& Neptune are in a 5:4 MMR (as labeled). Curves depicting the semi-major axes, perihelion, and apohelion of each planet are labeled. The current semi-major axes, perihelia and apohelia of the current solar system are plotted as grey points for comparison. In this model, the global instability is brought forth by Jupiter \& Uranus encountering a mutual 5:3 MMR.}
\end{figure}

\begin{figure}[t]
\includegraphics[width=0.5\textwidth]{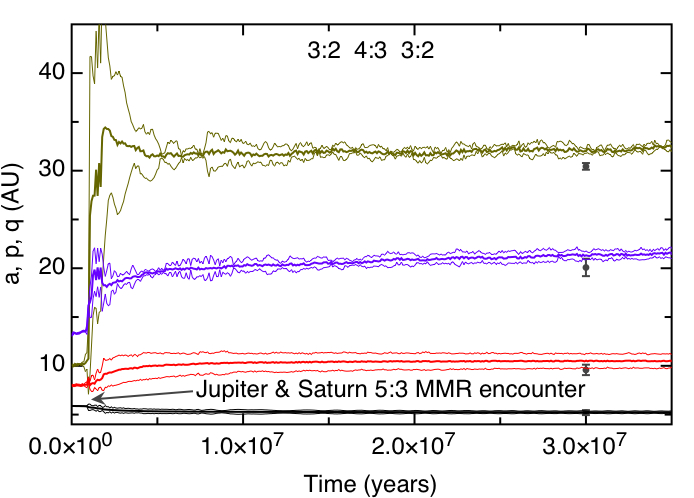}
\caption{Same as fig.1 except initially, Saturn \& Uranus are in a 4:3 MMR and Uranus \& Neptune are in a 3:2 MMR.}
\end{figure}

\begin{figure}[t]
\includegraphics[width=0.5\textwidth]{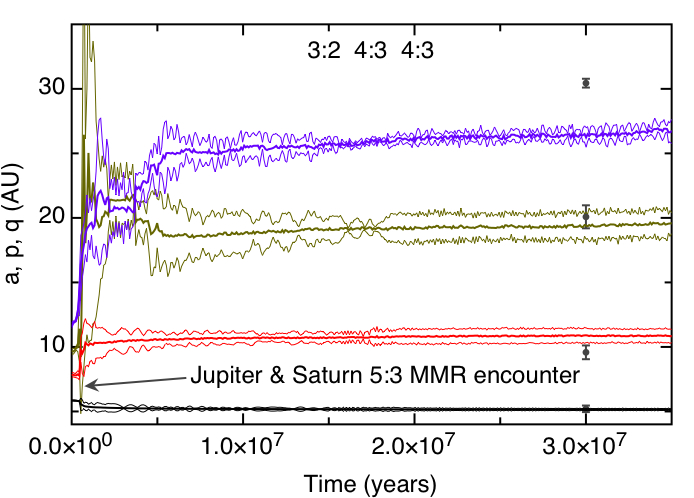}
\caption{Same as fig.1 except initially, both Saturn \& Uranus and Uranus \& Neptune are in 4:3 MMR's.}
\end{figure}

\begin{figure}[t]
\includegraphics[width=0.5\textwidth]{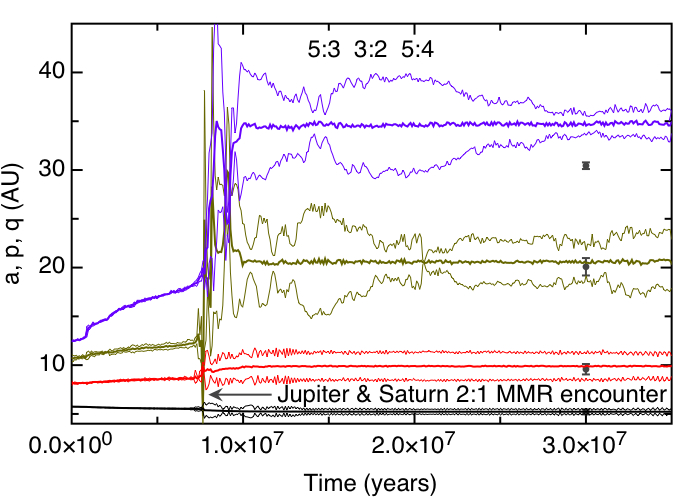}
\caption{Dynamical evolution of the initial configuration where initially Jupiter \& Saturn are in a 5:3 MMR, Saturn \& Uranus are  in a 3:2 MMR and Uranus \& Neptune are in a 5:4 MMR. In this model, the global instability is brought forth by Jupiter \& Saturn encountering a mutual 2:1 MMR, just as in the classical Nice model. All else is as in fig.1.}
\end{figure}

\subsection{Initial Conditions with Jupiter and Saturn in a 5:3 MMR}
We now move on to the next family of initial conditions. To begin with, we take the same approach as above. Stable multi-resonant configurations of this family are listed as the second set of entries in table (1). We simulated the evolutions of these systems with 20 integrations each. After completion, a clear boundary between initial conditions that result in smooth migration and those that result in scattering developed. Namely, all setups where Saturn \& Uranus are initially in a 2:1 MMR were characterized by smooth evolutions. A similar scenario describes the fate of initial conditions where Saturn \& Uranus are in a 3:2 MMR while Uranus \& Neptune are in a 3:2 or a 4:3 MMR. However in the same context, if Uranus and Neptune start out in a 5:4 or a 6:5 MMR, ice giant/gas giant scattering as well as transient phases of high eccentricities are present. Particularly, for the configuration where Uranus \& Neptune start out in a 5:4 MMR, 20\% of the integrations were successful with 50\% of them exhibiting close encounters between an ice giant and both gas giants. For the configuration where Uranus \& Neptune start out in a 6:5 MMR, also 20\% of the integrations were successful, but none of the solutions exhibited ice giant encounters with both gas giants.

In the subset of initial conditions where Saturn \& Uranus are in a 4:3 MMR, the configurations with Uranus \& Neptune in a 3:2 MMR and a 4:3 MMR can serve as good candidates for solar system formation, but the configuration with Uranus \& Neptune in a 5:4 MMR consistently leads to ejections. 10\% of the integrations with Uranus \& Neptune initially in a 3:2 MMR were successful, all of them exhibiting scattering with both ice giants. In the context of the configuration with Uranus \& Neptune initially in a 4:3 MMR, 15\% of the integrations were successful, while 33\% of them lead to encounters of an ice giant with both gas giants.

Examples of successful evolutions that start from the initial conditions described above are presented in figures (4) - (7), with final orbital parameters entered into table (2). Note that the scattering event in the evolution of the configuration where Saturn \& Uranus are initially in a 3:2 MMR and Uranus \& Neptune are in a 5:4 MMR (fig. 4) is considerably more violent than that in most other examples. This is because majority of close encounters here is between Jupiter and Uranus, while in most other simulations, Saturn is responsible for scattering. It is important to understand that this is not a unique feature of the particular initial condition. We have observed similar phenomena in simulations of other setups as well.

Note that in a scenario where Jupiter and Saturn start out in a 5:3 MMR, the instability is brought on by their crossing of the 2:1 MMR, just as in the classical Nice model. Much effort has been put into fine-tuning the classical Nice model's initial conditions (Tsiaganis et al. 2005, Morbidelli et al. 2005, Gomes et al. 2005, Levison et al. 2008). What matters most, however, are the locations of the planets when Jupiter and Saturn are crossing the 2:1 MMR. Let us now examine if the classical Nice model is compatible with any multi-resonant initial conditions from this family.

We begin our calculation by measuring the semi-major axes of the four planets at the Jupiter/Saturn 2:1 MMR crossing in figure (2a) of Levison et al. (2008). The values are listed in the second column of Table (3). Between encounters with MMR's, migrations of Jupiter and Saturn are mostly due to scattering of planetesimals. Malhotra (1995) showed from conservation of angular momentum that this process obeys a relation, which upon integration can be written as
\begin{equation}
\Delta\log{a} \simeq f \frac{m_{s}}{ m_{planet}}
\end{equation}
where $f$ is an empirically determined ``efficiency," listed in Table (3), and $m_{s}$ is the total scattered mass. When applied to Jupiter and Saturn simultaneously, this relation can be used to ``backtrace" the system's migration, by roughly estimating the starting semi-major axes of Jupiter and the scattered mass $m_{s}$. Setting $\Delta a_S = (5/3)^{2/3} a_J^i - (2)^{2/3} a_J^f$ and $\Delta a_J = a_J^f - a_J^i$ with $a_J^f = 5.45$ AU yields a total scattered mass of $37 m_{\oplus}$ and $a_J^i = 5.69$ AU. 

Neglecting high-order resonant encounters, we then use the calculated scattered mass and apply equation (4) to Uranus and Neptune to determine their original positions. The back-traced initial conditions are listed in the third column of table (3). Incidentally, these initial conditions are close to a multi-resonant configuration where Saturn \& Uranus and Uranus \& Neptune are both in 4:3 MMR's. Recall that this initial condition is indeed one of the setups that consistently exhibit scattering. However, given the similarities in dynamical evolutions among the successful initial conditions of this family, at this level of accuracy, it is probably safe to say that all four of them are compatible with the classical Nice model results.

\begin{figure}[t]
\includegraphics[width=0.5\textwidth]{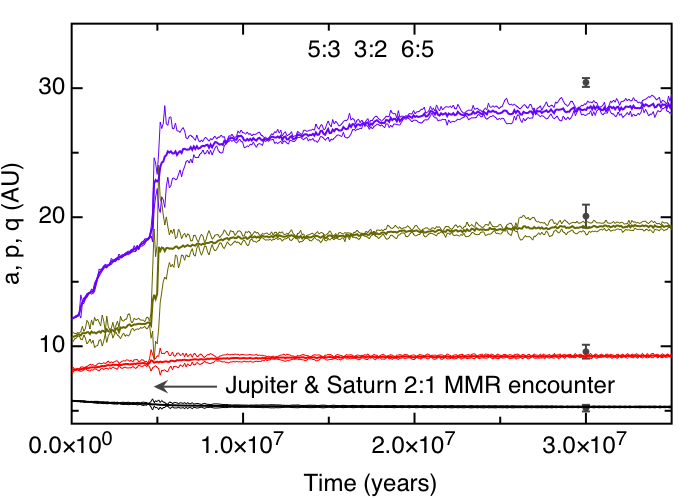}
\caption{Same as fig.4 except initially, Saturn \& Uranus are in a 3:2 MMR and Uranus \& Neptune are in a 6:5 MMR.}
\end{figure}

\begin{figure}[t]
\includegraphics[width=0.5\textwidth]{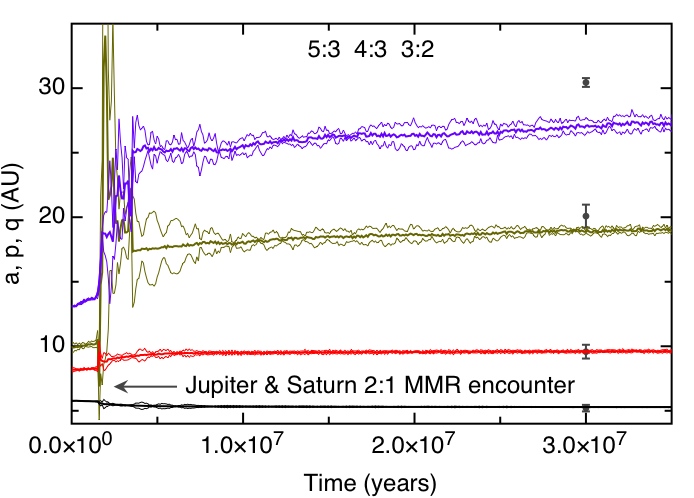}
\caption{Same as fig.4 except initially, Saturn \& Uranus are in a 4:3 MMR and Uranus \& Neptune are in a 3:2 MMR.}
\end{figure}

\begin{figure}[t]
\includegraphics[width=0.5\textwidth]{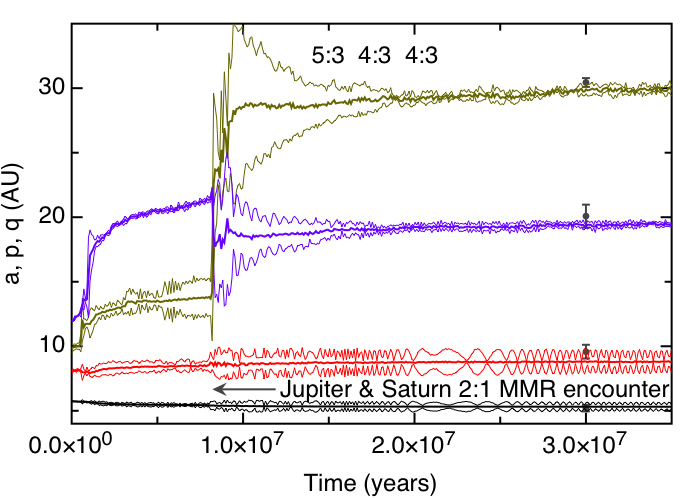}
\caption{Same as fig.4 except initially, both Saturn \& Uranus and Uranus \& Neptune are in 4:3 MMR's.}
\end{figure}

\subsection{Initial Conditions with Jupiter and Saturn in a 2:1 MMR}
Let us now consider the final family of initial conditions, listed in table (1), where Jupiter and Saturn are initially in a 2:1 MMR. Unlike the scenario of the classical Nice model (Tsiaganis et al. 2005), there are no major resonances to cross for Jupiter and Saturn between the 2:1 and the 5:2 MMR's. Consequently, a different mechanism, involving different resonances, is needed to create the instability. Thommes et al. (2008) considered the dynamical evolution of a system where Jupiter \& Saturn are in a 2:1 MMR, Saturn \& Uranus are in a 3:2 MMR, and Uranus \& Neptune are in a 4:3 MMR. In such a system, the instability is triggered by Uranus and Neptune crossing a 7:5 MMR. Due to a weaker, second-order nature of this resonance, the eccentricity increase is rather small. Incidentally in this particular system, this is enough for the ice giants to cross orbits and scatter off of each other, but not off of one of the gas giants\footnote{Another scenario present in the integrations of Thommes et al. (2008) is the escape of one of the ice giants. In this case, the remaining ice giant is left with a high eccentricity, but there are only 3 planets left in the system.}. It appears that somewhat larger eccentricities are needed. Testing each initial condition with a large number of numerical simulations, as discussed above, is rather time-consuming. Consequently, it is worthwhile to quantify the amplitudes of eccentricity jumps due to various resonance crossings before-hand if possible. For this set of initial conditions, under the assumption of adiabatic migration, the eccentricity jumps are deterministic and can be estimated analytically (Henrard 1982).

Following the treatment of (Peale 1986, see also Murray \& Dermott 1999), we consider the planar internal first-order $j:(j-1)$ resonant Hamiltonian
\begin{eqnarray}
\mathit{H}_{res} &=& - \frac{G^2 m_{\odot}^2 m^3}{2 \Lambda^2} - \frac{G^2 m_{\odot}^2 m'^3}{2 \Lambda'^2} \nonumber \\ 
&-& \frac{G^2 m_{\odot}^2 m m'^3}{\Lambda'^2} f(a/a')
\sqrt{\frac{2 \Gamma}{\Lambda}} \cos(j \lambda' + (1-j)\lambda + \gamma ) \nonumber \\
&-& \Gamma \dot{\gamma}_{sec} + \Lambda \dot{\lambda}_{sec} - \Gamma' \dot{\gamma}'_{sec} + \Lambda' \dot{\lambda}'_{sec}
\end{eqnarray}
where $\lambda$ is the mean longitude, $\gamma=\varpi$ is the longitude of perihelion, $\Lambda = (m \ m_{\odot})/(m + m_{\odot})\sqrt{G(m_{\odot}+m)a}$ \& $\Gamma = \Lambda (1-\sqrt{1-e^2})$ are their respective Poincar\'e conjugate momenta, and the prime designates the outer planet. The secular changes in mean longitude and longitude of perihelion are accounted for by the last four terms, while $f(a/a')$ arises from the classical expansion of the planetary disturbing potential and is a function of Laplace coefficients and their derivatives. The expressions for $f(a/a')$ are presented in Appendix B of (Murray \& Dermott 1999). Under a series of variable transformations (see Peale 1986 for derivation), this Hamiltonian can be rewritten to take a simpler form. Let us introduce the constants $\alpha$, $\beta$ and $\epsilon$:
\begin{equation}
\alpha = (j-1) n^{*} - j  n'^{*} + \dot{\gamma}_{sec} ,
\end{equation}
\begin{equation}
\beta = \frac{3}{2} \left[ \frac{ (j-1)^2 }{m a^2} + \frac{j^2}{m' a'^2}   \right] ,
\end{equation}
\begin{equation}
\epsilon = n^{3/2} f(a/a') \frac{a^2}{a'}   \frac{m'}{m_{\odot}}   \sqrt{m},
\end{equation}
where $n^*$ is sum of the Keplerian mean motion and the secular change in mean longitudes. It is important to note that these expressions are not strictly constant, since semi-major axis changes. However, $\beta$ is only weakly dependent on semi-major axis (Peale 1976), and in the case of $\epsilon$, variations due to the cosine term dominate, so the assumption of constant coefficients is sound (Murray and Dermott 1999). Relative to the original Hamiltonian, we scale the momentum as:
\begin{equation}
\Phi=\Gamma \left( \frac{2 \beta}{\epsilon} \right)^{2/3}.
\end{equation}
The corresponding conjugate angle $\phi$ is simply the cosine argument in equation (5), although if $\epsilon > 0$, we also need to add $\pi$ to the expression (Murray \& Dermott 1999). That said, the transformed Hamiltonian takes the form
\begin{equation}
H_{res} = \delta \Phi + \Phi^2 - 2 \sqrt{2\Phi} \cos(\phi).
\end{equation}
This Hamiltonian is parameterized by
\begin{equation}
\delta = \alpha \left( \frac{4	}{\epsilon^2 \beta}   \right)^{1/3} ,
\end{equation}
which is a measure of the perturbed object's proximity to exact resonance. Finally we note that this Hamiltonian is most easily visualized in terms of polar coordinates, so we introduce the new mixed canonical variables $x = \sqrt{2 \Phi} \cos{\phi}$ and $y = \sqrt{2 \Phi} \sin{\phi}$ (Henrard 
\& Lamaitre 1983). The Hamiltonian now takes the form
\begin{equation}
H_{res} = \frac{\delta (x^2 + y^2)}{2} + \frac{(x^2 + y^2)^2}{4} -2x.
\end{equation}
Upon application of Hamilton's equations, we see that the stationary points of the above Hamiltonian are described by the equation
\begin{equation}
x^3 + \delta x -2 =0.
\end{equation}
For resonant encounters aided by divergent migration, $\delta < 3$ initially. In this case, the existence of a separatrix is ensured, and there are three real fixed points, all of which lie on the x-axis. Two of these points are always negative, and the more negative one is unstable, as it lies on the intersection of the inner and the outer branches of the separatrix. This is crucial to the estimation of eccentricity jumps during resonant encounters. 

If migration is slow enough for $\delta$ to be approximately constant over one period of motion, the action, defined as
\begin{equation}
J = \oint \Phi d \phi = \oint x dy
\end{equation}
is an adiabatic invariant (Peale 1986). In other words, it is constant except during separatrix crossing. Furthermore, when the separatrix is far away, the trajectories of the circulating orbits in ($x,y$) space are circles to a good approximation. Consequently, we can write $J = 2 \pi \Phi$ (Murray \& Dermott 1999).

When two planets approach commensurability, a wide separatrix is seen as shrinking down on the orbit of the perturbed planet in ($x,y$) space. When the inner branch of the separatrix engulfs the planetary orbit, the process of resonance crossing is characterized by the planet switching to the separatrix's outer circulating branch. The outer branch has a wider radius, thus the increase in action. However, during this switch, the perturbed planet must necessarily pass through the unstable stationary point described above. Consequently, the calculation is as follows: knowing the action prior to the resonant encounter, we can determine the value of $\delta$ at the transition using equation (13). Recall however, that $\delta$ also parameterizes the Hamiltonian, and therefore determines the shape of the separatrix, while the area engulfed by the outer branch corresponds to the new action (see supplemental material of Tsiaganis et al. 2005 for an intuitive discussion). It can be shown that the actions before and after resonance crossing are related by 
\begin{equation}
J_{i}+J_{f} = - 2 \pi \delta.
\end{equation}
Thus, the new eccentricity can be easily backed out. 

The above analysis can also be applied to external resonances. In this case, $\gamma$ in the cosine argument of the Hamiltonian (5) is replaced by $\gamma'$, and its factor $\sqrt{2 \Gamma / \Lambda}$ is replaced by $\sqrt{2 \Gamma' / \Lambda'}$, since we are now concerned with an $e'$ resonance. Accordingly, we change the scaling factors to
\begin{equation}
\alpha = (j-1) n^{*} - j  n'^{*} + \dot{\gamma}'_{sec} ,
\end{equation}
\begin{equation}
\epsilon = n'^{3/2} f(a/a') a'   \frac{m}{m_{\odot}}   \sqrt{m'},
\end{equation}
while $\beta$ remains the same. Note also that any indirect terms in the expansion of the disturbing function must be accounted for in $f(a/a')$. Under these transformations, Hamiltonian (10) still applies, and so does the subsequent analysis (Murray and Dermott 1999).

The resulting estimates of eccentricity jumps for various first-order resonant encounters between Uranus \& Neptune and Saturn \& Uranus are listed in Table (4). As can be seen from these calculations, all first-order resonant encounters between Uranus and Neptune produce rather small eccentricity jumps. Therefore, we disfavor them as good options for triggering instability scenarios in which encounters with Saturn take place. Numerical integrations performed in the two previous sections are suggestive of this as well. We therefore rule out the configurations where Saturn and Uranus are in a 2:1 MMR. Resonant encounters between Saturn and Uranus, however, are a different story: in all cases, Uranus acquires an eccentricity comparable to $0.1$. Simulations reveal that the configurations where Saturn and Uranus are in a 3:2 MMR do not result in strong instabilities. This is because the system is given a chance to encounter high-order MMR's between Uranus and Neptune and spread out before crossing the 2:1 MMR between Saturn \& Uranus. Furthermore, in order to increase the chances of Uranus/Saturn orbital crossing, it helps to start the two planets in the most compact stable resonance - namely the 4:3. From here, the degeneracy lies in whether Uranus and Neptune start out in a 3:2 or 4:3 MMR. Out of 30 numerical simulations performed for each configuration, we only observed ice-giant/gas-giant scattering events in the evolutions of the system where Uranus and Neptune are in the 4:3 MMR. Particularly, 23\% were successful, and in 57\% of the successful integrations, an ice giant exhibited encounters with both gas giants. Figure (8) shows the dynamical evolution of this configuration with time, while table (2) lists the final values of semi-major axes, eccentricities, and inclinations for the planets. 

Aside from Saturn's close encounter with an ice-giant, Jupiter's and Saturn's migration is dominated by scattering of planetesimals. As a result, equation (4) approximates their evolution well. Similarly to the previous section, when applied to Jupiter and Saturn simultaneously, this relation can be used to roughly estimate the starting semi-major axis of Jupiter and the initial $m_{disk}$. Setting $\Delta a_S = (2)^{2/3} a_J^i - (5/2)^{2/3} a_J^f$ and $\Delta a_J = a_J^f - a_J^i$ with $a_J^f = 5.2$ AU yields a total scattered mass of $49 m_{\oplus}$ and $a_J^i = 5.5$ AU. These values are in good agreement with numerical integrations. For instance, the evolution presented in figure (1) started had the parameters: $a_J^i = 5.5$ AU and $m_{disk} = 51m_{\oplus}$.

\begin{figure}[t]
\includegraphics[width=0.5\textwidth]{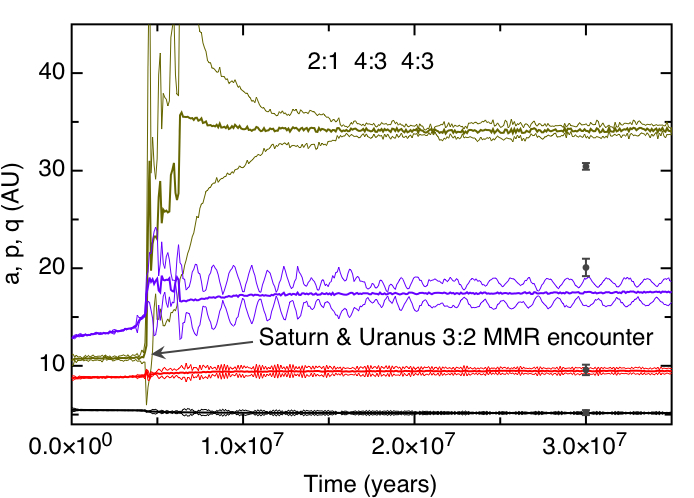}
\caption{Dynamical evolution of the initial configuration where initially Jupiter \& Saturn are in a 2:1 MMR, while both Saturn \& Uranus and Uranus \& Neptune are in a 4:3 MMR's. In this model, the global instability is brought forth by Saturn \& Uranus encountering a mutual 3:2 MMR. All else is as in fig.1.}
\end{figure}

\section{Discussion} 

The calculations presented in this work aim to place constraints on the early dynamical state of the solar system. We began by constructing a large array of multi-resonant systems. We then considered each one of these systems individually and tested them against two inter-related constraints: scattering of an ice-giant from a gas-giant and the outer ice giant undergoing a transient phase of high eccentricity. Both events are needed to reproduce the current dynamical architecture of the solar system. 

We showed numerically that three configurations in which Jupiter \& Saturn start out in a 3:2 MMR follow evolutionary tracks which are compatible with our constraints. The three systems are those in which (1) Saturn \& Uranus are in a 3:2 MMR while Uranus \& Neptune are in a 5:4 MMR, (2) Saturn \& Uranus are in a 4:3 MMR while Uranus \& Neptune are in a 3:2 MMR and (3) Saturn \& Uranus are in a 4:3 MMR while Uranus \& Neptune are also in a 4:3 MMR. Two of these configurations were previously thought to be unstable (Morbidelli et al. 2007).

We showed that the classic Nice model can be reproduced from multi-resonant initial conditions where Juipter \& Saturn start out in a 5:3 MMR. The four particular primordial states that we found were those where initially, (1) Saturn \& Uranus are in a 3:2 MMR while Uranus \& Neptune are in a 5:4 MMR, (2) Saturn \& Uranus are in a 3:2 MMR while Uranus \& Neptune are in a 6:5 MMR, (3) Saturn \& Uranus are in a 4:3 MMR while Uranus \& Neptune are in a 3:2 MMR, and (4) Saturn \& Uranus are initially in a 4:3 MMR while Uranus \& Neptune are also in a 4:3 MMR.

Finally, we used an analytical technique to rule out a large portion of the generated multi-resonant systems, in which Jupiter and Saturn are initially in the 2:1 MMR, based on an argument that the eccentricities generated by any Uranus/Neptune resonant encounters are too small. Simultaneously, we showed that the considered constraints can be satisfied by the dynamical evolution whose initial condition has Jupiter \& Saturn locked in a 2:1 MMR, and the other pairs of the planets in 4:3 MMR's.

The calculations presented here are intended in part as a point of departure for future research. There is indeed a large array of unexplored issues. For instance, in the case of the system where Jupiter \& Saturn start out in a 2:1 MMR, it is not clear if a single scattering event alone is enough to correctly reproduce the secular dynamics of Jupiter and Saturn, or if a resonant encounter is also required. In the initial condition where Jupiter and Saturn start out in a 5:3 MMR, dynamical stability may pose an issue (Morbidelli \& Crida 2007). A further criterion of interest is LHB. While we do not attempt to time the onset of instability with LHB in these simulations, it is certainly fair to ask if the dynamical evolutions presented here are compatible with a long quiescent period preceding any resonant encounters. Two related issues immediately follow. First, will changing the placement of the inner boundary of the planetesimal swarm qualitatively change the process responsible for the onset of the instability of the dynamical evolutions? Second, how will the effective viscosity that arises from the self-gravity of the disk affect our results? A large-scale numerical modeling effort will be instrumental in providing these answers. 

There is certainly room for broader study of the current setup of the problem as well. In this work, we have restricted ourselves to four-planet multi-resonant configurations. Certainly, the idea of initially forming more than two ice giants is not unreasonable (Ford \& Chiang 2007, see however Levison \& Morbidelli 2007). Although the results of Morbidelli et al. (2007) suggests that additional planets in a compact multi-resonant system compromise dynamical stability, more work is needed to obtain a good handle on this part of the problem.
 
In conclusion, the determination of a small subset of initial conditions allows for a much more efficient survey of the parameter space. In this work, we have taken a step in this direction. We must keep in mind that the system at hand is highly chaotic, and must in the end be studied numerically. The resulting determinations are often probabilistic rather than conclusive, however the results are certainly bound to gain statistical weight as the number of completed simulations increases. Thus, while much progress is yet to be made, additional research carries great value since a solid understanding of initial conditions plays an unavoidably important role in further development of a comprehensive model for solar system's formation. 

\acknowledgments We thank Hal Levison, Alessandro Morbidelli, Ramon Brasser, Gregory Laughlin and Darin Ragozzine for useful discussions.

\begin{table}
\begin{center}
\caption{Orbital Elements at the End of Simulations}
\begin{tabular}{lllllllcccc}
\\
\tableline\tableline
  & $a$ (AU)  & $e$ & $i$ (deg.)\\
\tableline

3:2 \ 3:2 \ 5:4 \ configuration  \ \ \ \ \ $k = 1.93$ \ \ \ \ \ $m_{disk} = 91 M_{\oplus}$\\ 
\tableline
Jupiter & 5.2 & 0.013 & 0.17 \\
Saturn & 11.2 & 0.025 & 0.18 \\
Uranus & 19.2 & 0.017 & 0.9 \\
Neptune & 31.5 & 0.018 & 1.3 \\
\tableline
3:2 \ 4:3 \ 3:2 \ configuration$^{\dagger}$  \ \ \ \ \ $k = 1.0$ \ \ \ \ \ $m_{disk} = 82 M_{\oplus}$\\ 
\tableline
Jupiter & 5.2 & 0.027 & 0.31 \\
Saturn & 10.5 & 0.068 & 0.5 \\
Uranus & 21.5 & 0.022 & 0.9 \\
Neptune & 32.5 & 0.011 & 0.85 \\
\tableline
3:2 \ 4:3 \ 4:3 \ configuration$^{\dagger}$ \ \ \ \ \ $k = 1.41$ \ \ \ \ \ $m_{disk} = 75 M_{\oplus}$ \\ 
\tableline
Jupiter & 5.15 & 0.018 & 0.55 \\
Saturn & 10.8 & 0.05 & 1.15 \\
Uranus & 19.6 & 0.036 & 1.63 \\
Neptune & 26.7 & 0.043 & 2.95 \\

\tableline
5:3 \ 3:2 \ 5:4 \ configuration  \ \ \ \ \ $k = 1.50$ \ \ \ \ \ $m_{disk} = 60 M_{\oplus}$ \\ 
\tableline
Jupiter & 5.22 & 0.073 & 0.37 \\
Saturn & 9.9 & 0.109 & 1.24 \\
Uranus & 20.39 & 0.122 & 2.66 \\
Neptune & 34.89 & 0.034 & 0.65 \\
\tableline
5:3 \ 3:2 \ 6:5 \ configuration  \ \ \ \ \ $k = 1.35$ \ \ \ \ \ $m_{disk} = 63 M_{\oplus}$ \\ 
\tableline
Jupiter & 5.3 & 0.011 & 0.07 \\
Saturn & 9.28 & 0.016 & 0.27 \\
Uranus & 19.23 & 0.008 & 0.08 \\
Neptune & 28.51 & 0.022 & 0.57 \\
\tableline
5:3 \ 4:3 \ 3:2 \ configuration  \ \ \ \ \ $k = 1.85$ \ \ \ \ \ $m_{disk} = 64 M_{\oplus}$ \\ 
\tableline
Jupiter & 5.29 & 0.004 & 0.49 \\
Saturn & 9.64 & 0.013 & 1.5 \\
Uranus & 18.99 & 0.016 & 0.69 \\
Neptune & 27.38 & 0.022 & 0.22 \\
\tableline
5:3 \ 4:3 \ 4:3 \ configuration$^{\dagger}$  \ \ \ \ \ $k = 1.75$ \ \ \ \ \ $m_{disk} = 58 M_{\oplus}$ \\ 
\tableline
Jupiter & 5.3 & 0.02 & 0.25 \\
Saturn & 8.8 & 0.09 & 0.14 \\
Neptune & 19.7 & 0.01 & 0.65 \\
Neptune & 30.4 & 0.007 & 1.86 \\

\tableline
2:1 \ 4:3 \ 4:3 \ configuration$^{\dagger}$  \ \ \ \ \ $k = 1.9$ \ \ \ \ \ $m_{disk} = 51 M_{\oplus}$\\ 
\tableline
Jupiter & 5.16 & 0.016 & 0.08 \\
Saturn & 9.48 & 0.029 & 0.13 \\
Uranus & 17.57 & 0.06 & 0.76 \\
Neptune & 34.34 & 0.004 & 0.6 \\

\tableline
\tableline
\end{tabular}
\end{center}
\tablenotetext{0}{Orbital elements of solar system analogues, resulting from different initial conditions, at the end of the dynamical evolution simulations, presented in figures 1 - 8. Simulations where the ice giants switched places are labeled with a ${\dagger}$. The disk mass used in each simulation, as well as the disk's power law index $k$ are also given. }
\end{table}

\begin{table}
\begin{center}
\caption{Analytical Calculation of Planetesimal-Driven Migration}
\begin{tabular}{llllllllcc}
\\
\tableline\tableline
Planet & $ f$ & $a$ (AU) (J:S 2:1)  & $a$ (AU) (J:S 5:3) & $a$ (AU) (5:4 4:3 4:3) \\
\tableline
Jupiter & 0.35 & 5.45 & 5.69 & 5.7  \\
Saturn & 0.2 & 8.74 & 7.97 & 8.02  \\
Uranus & 0.08 & 12.61 & 9.84 & 9.71  \\
Jupiter & 0.15 & 18.36 & 12.34 & 11.77  \\
\tableline
\end{tabular}
\end{center}
\tablenotetext{0}{Inputs and results of the analytical calculations of planetesimal-driven migration (section 3.2). $f$ is the effective scattering efficiency, inferred from numerical simulations. Although it varies from run to run, the variation is not too great. The second column, $a$ (AU) (J:S 2:1), lists the positions of the planets at the time of Jupiter/Saturn 2:1 MMR crossing, as inferred from the results of Levison et al. (2008). The next column,  $a$ (AU) (J:S 5:3), list the semi-major axes of the planets with Jupiter \& Saturn nominally in the 5:3 MMR, traced back using equation (4) from the previous column. The last column, $a$ (AU) (5:4 4:3 4:3), lists the semi-major axes of the 5:4 4:3 4:3 multi-resonant configuration, assembled as discussed in section 2. Note the quantitative similarity between the two right-most columns. This leads us to believe that the 5:4 4:3 4:3 multi-resonant configuration, as well as other compact configurations from the same family are compatible with the classical Nice model. }
\end{table}

\begin{table}
\begin{center}
\caption{Analytical Estimates of Eccentricities After a Resonant Encounter}
\begin{tabular}{cccccccccc}
\\
\tableline\tableline
Resonance & $e_N$  & $e_U$\\
\tableline
3:2 & 0.031 & 0.037\\
4:3 & 0.027 & 0.031\\
5:4 & 0.024 & 0.028\\
\tableline
Resonance & $e_U$  & $e_S$\\
\tableline
2:1 & 0.062 & 0.022\\
3:2 & 0.098 & 0.018\\
4:3 & 0.084 & 0.016\\
5:4 & 0.075 & 0.015\\
\tableline
\end{tabular}
\end{center}
\end{table}

\end{document}